
\def\cH{{\cal H}}
\bigskip
\bigskip
\centerline{PLATEAU OF THE MAGNETIZATION CURVE OF THE $S = 1/2$}
\bigskip
\centerline{FERROMAGNETIC-FERROMAGNETIC-ANTIFERROMAGNETIC}
\bigskip
\centerline{SPIN CHAIN}
\bigskip
\bigskip
\centerline{Kiyomi Okamoto}
\bigskip
\centerline{Department of Physics, Tokyo Institute of Technology,
Meguro, Tokyo 152, Japan}
\bigskip
\centerline{(Received~~~~~~~~~~~~~~~~~~~~~~~~)}

\bigskip
\bigskip
\bigskip
I analytically study the plateau of the magnetization curve at
$M/M_{\rm S} = 1/3$ (where $M_{\rm S}$ is the saturation magnetization)
of the one-dimensional $S=1/2$ trimerized Heisenberg spin system with
ferromagnetic ($J_{\rm F}$)-ferromagnetic ($J_{\rm
F}$)-antiferromagnetic ($J_{\rm A}$) interactions at $T=0$.
I use the bosonization technique for the fermion representation of the
spin Hamiltonian through the Jordan-Wigner transformation.
The plateau appears when $\gamma \equiv J_{\rm F}/J_{\rm A}
\allowbreak < \gamma_{\rm C}$, and vanishes when $\gamma > \gamma_{\rm
C}$, where the critical value $\gamma_{\rm C}$ is estimated as $\gamma_{\rm C}
= 5 \sim 6$.
The behavior of the width of the plateau near $\gamma_{\rm C}$ is of
the Kosterlitz-Thouless type.
The present theory well explains the numerical result by Hida.

\vfill\eject

{}~~~~~~~~Recently Hida${}^1$ numerically studied the $S=1/2$ spin chain
system in which the interaction between neighboring spins changes as
ferromagnetic-ferromag-netic-antiferromagnetic (Fig.1).
The Hamiltonian of this model is written by
$$
  \eqalignno{
    &\cH
    =   J_{\rm A} \sum_j
                   (h^\perp_{3j-1,3j} + \Delta h^z_{3j-1,3j})  \cr
    &~~~~~~~- J_{\rm F} \sum_j
              (h^\perp_{3j,3j+1} + \Delta h^z_{3j,3j+1}  \cr
    &~~~~~~~~~~~~~~~~~~~~
              + h^\perp_{3j+1,3j+2} + \Delta h^z_{3j+1,3j+2})~,~~~
    &(1)  \cr
    &h^\perp_{l,m}
    \equiv S_l^x S_m^x + S_l^y S_m^y~,~~~~
    h^z_{l,m} \equiv S_l^z S_m^z~,
    &(2)  }
$$
where $J_{\rm A}>0$ and $J_{\rm F}>0$ are the magnitudes of the
antiferromagnetic and ferromagnetic couplings, respectively.
I have introduced the anisotropy parameter $\Delta>0$ for later
convenience, although Hida investigated only the $\Delta = 1$ case.
This model is not too artificial nor a theoretical toy, as noticed by
Hida.
In fact, the substance ${\rm 3CuCl_2 \cdot 2dioxane}$, of which
magnetization is measured by Ajiro {\it et al.}${}^2$ in strong
magnetic fields, is known to be a (quasi-)one dimensional magnet to
which this model is applied.

\bigskip

{}~~~~~~~~Hida${}^1$ investigated this model (only the $\Delta = 1$ case)
by the numerical diagonalization method for finite systems (up to 24
spins) to find that there was a plateau in the magnetization curve as
far as $\gamma \equiv J_{\rm F}/J_{\rm A}$ is small, and the width of
the plateau decreases as the parameter $\gamma$ increases (Fig.2).
The location of the plateau was $M_{\rm S}/3$, where $M_{\rm S}$ is
the saturation magnetization.
He could not obtain a definite conclusion about the existence of the
plateau for large $\gamma$ case, because the magnetization varied
stepwise for the finite systems.
Since this model becomes $S=3/2$ antiferromagnetic chain model when
$\Delta = 1$ and $\gamma \rightarrow \infty$ due to the formation of
an $S=3/2$ quartet by spins ${\bf S}_{3j}$, ${\bf S}_{3j+1}$ and ${\bf
S}_{3j+2}$, it is expected that there is no plateau in this limit.
It is believed that all the half-odd spin chains with
antiferromagnetic nearest-neighbor interactions belong to the same
universality class as that of $S=1/2$.

\bigskip

{}~~~~~~~~There arise several questions:

(i) Why does the plateau appear at $M/M_{\rm S} = 1/3$?
(ii) Does the plateau vanish at a finite value of $\gamma$ or persist
to $\gamma = \infty$?
(iii) If the plateau vanishes at $\gamma_{\rm C}$, what is the
behavior of the plateau near $\gamma_{\rm C}$?

They are interesting questions not only from the standpoint of the
statistical physics but also from that of explaining the properties of
the existing materials.
I note that the plateau was not observed in the report of Ajiro {\it
et al}.${}^2$

\bigskip

{}~~~~~~~~In this paper I will analytically study the mechanism for the
appearance of the plateau and also discuss whether the critical value
$\gamma_{\rm C}$ exists or not.

\bigskip

{}~~~~~~~~Performing the spin rotation around the $z$-axis for the spins
located at the sites $3j$
$$
    S_{3j}^x \Rightarrow -S_{3j}^x~,~~
    S_{3j}^y \Rightarrow -S_{3j}^y~,~~
    S_{3j}^z \Rightarrow  S_{3j}^z~,
    \eqno(3)
$$
Hamiltonian (1) is transformed into the form of a generalized version
of the trimerized antiferromagne-tic chain:
$$
  \eqalign{
    \cH =
    &J_0(1 - 2\delta_\perp) \sum_j h^\perp_{3j-1,3j}  \cr
    & + J_0(1 + \delta_\perp)  \sum_j (h^\perp_{3j,3j+1}
                                    + h^\perp_{3j+1,3j+2})  \cr
    &+ J_0(\Delta_0 - 2\delta_z) \sum_j h^z_{3j-1,3j}  \cr
    &+ J_0(\Delta_0 + \delta_z) \sum_j (h^z_{3j,3j+1}
                                      + h^z_{3j+1,3j+2})~, }
    \eqno(4)
$$
where
$$
  \eqalign{
    &J_0 = {2J_{\rm F} + J_{\rm A} \over 3}~,~~~
    \Delta_0 = {-2\gamma + 1 \over 2\gamma + 1}\,\Delta~,\cr
    &\delta_\perp = {\gamma - 1 \over 2\gamma + 1}~,~~~
    \delta_z = - \,{\gamma + 1 \over 2\gamma + 1}\,\Delta~.}
    \eqno(5)
$$

\bigskip

{}~~~~~In case of the $XY$ model where $\Delta = 0$, we can solve
Hamiltonian (4) by the use of the fermion representation through the
Jordan-Wigner transformation.
The dispersion relation is shown in Fig.3.
When the magnetic field is applied, the dispersion curve shifts along
the $\omega$-axis, which explains the existence of the plateau as well
as its location $M/M_{\rm S} = 1/3$.
Note that the magnetization $M$ is related to the number of occupied
states as
$$
      {M \over N}
      = {1 \over 2} - {{\rm (occupied~states)} \over N}~,
      \eqno(6)
$$
where $N$ is the total number of spins.

\bigskip

{}~~~~~This is a simple explanation for the question why the plateau
appears at $M/M_{\rm S} = 1/3$.
However, there still remains questions.
Why is there no plateau in the experimental results on ${\rm 3CuCl_2
\cdot 2dioxane}$?
Is there any essential difference between the $XY$ case ($\Delta = 0$)
and the isotropic Heisenberg case ($\Delta = 1$)?
How can we explain Hida's result, especially in $\gamma \ge 5$ case?
To answer these questions, we have to consider the interactions
between fermions when $\Delta \ne 0$.

\bigskip

{}~~~~~~~~~To consider the effect of the trimerization and that of the
interaction between fermions simultaneously, I use the method of the
bosonization.${}^3$
The bosonization is one of the powerful methods in one-dimensional
quantum problems.
Here I do not enter into the details of the bosonization procedure.
I note that the bosonization is usually done in case of no
magnetization (half-filled in the language of the fermion), but in the
present case, we have to perform the bosonization near $M/M_{\rm S} =
1/3$ (1/3-filled in the language of the fermion).

\bigskip

{}~~~~~After the bosonization, Hamiltonian (3) is transformed into a
generalized sine-Gordon Hamiltonian:
$$
  \eqalignno{
    &\tilde\cH
    = {\sqrt{3}J_0 a \over 2} \int dx
      \{ A(\nabla\theta)^2 + CP^2  \cr
    &~~~~~~~~~~- B_\perp \cos\theta
               - B_z (\nabla\theta)^2 \cos\theta \}~,
    &(7)  \cr
    &A
    = {1 \over 8\pi}
      \left(1 + {2\sqrt{3} \Delta_0 \over \pi} \right),~
    C
    = 2\pi \left( 1 - {2\Delta_0 \over \sqrt{3}\pi} \right),
    &(8)  \cr
    &B_\perp
    = 2\delta_\perp / \sqrt{3}a^2~,~~~~
    B_z
    = \delta_z / \pi~,
    &(9)  \cr
    &[\theta(x),~P(x')] = {\rm i}\delta(x-x')~,
    &(10)  }
$$
where $a$ is the distance between neighboring spins.
The effect of the trimerization appears in the $B_\perp$ and $B_z$
terms.
If the $B_\perp$ term and/or the $B_z$ term are relevant in the sense
of the renormalization group, the spectrum of $\tilde\cH$ has a gap,
which brings about the plateau in the magnetization curve.
If both of them are irrelevant, the spectrum of $\tilde\cH$ is gapless,
which results in no plateau.
Which case is realized? --- It depends on the magnitudes of $B_\perp$
and $B_z$ and also on the parameter
$$
    \eta \equiv {2\pi}^{-1} \sqrt{C / A}~.
    \eqno(11)
$$
The renormalization group calculation shows that the $B_\perp$ term
and/or the $B_z$ term are relevant as far as $\eta<4$.

\bigskip

{}~~~~~~~~The value of $\eta$ is slightly shifted through the
bosonization procedure.
The expressions of $A$ and $B$ in eq.(8) is considered to be the
lowest order expansions with respect to $\Delta_0$.
When $\Delta_0 \rightarrow -1$ (i.e. $\gamma \rightarrow \infty$), the
system becomes ferromagnetic.
Therefore $\eta$ should diverge to $+\infty$ when $\Delta_0
\rightarrow -1$, although it seems to diverge at $\Delta_0 \rightarrow
-\pi/2\sqrt{3} =  -0.907$ from eq.(8).

\bigskip

{}~~~~~~~~The bosonized Hamiltonian (7) has the same form as that of the
generalized version of the dimerized $XXZ$ model.
$$
  \eqalign{
    \cH^{({\rm d})}
    = &J \sum_l \{1 + (-1)^l \delta_\perp^{({\rm d})}\} h^\perp_{l,l+1}
       \cr
      &+ J \sum_l \{\Delta^{({\rm d})} + (-1)^l \delta_z^{({\rm d})}\}
        h^z_{l,l+1}~,
    ~~~~J>0~,}
    \eqno(12)
$$
In fact, if we perform the bosonization for Hamiltonian (12), we
obtain${}^4$
$$
  \eqalign{
    \tilde\cH^{({\rm d})} =
    &Ja \int dx \{ A^{({\rm d})}(\nabla\theta)^2 + C^{({\rm d})}P^2
\cr
    &~~~~~   - B_\perp^{({\rm d})} \cos\theta
              - B_z^{({\rm d})} (\nabla\theta)^2 \cos\theta \}~,}
    \eqno(13)
$$
where
$$
  \eqalignno{
    &A^{({\rm d})}
    = {a \over 8\pi}
      \left(1 + {3\Delta^{({\rm d})} \over \pi} \right)~,~~~
    C^{({\rm d})}
    = 2\pi a
      \left( 1 - {\Delta^{({\rm d})} \over \pi} \right)~.
    &(14)  \cr
    &B_\perp^{({\rm d})}
    = \delta_\perp^{({\rm d})} / a^2~,~~~~~
    B_z^{({\rm d})}
    = \delta_z^{({\rm d})} / \pi~.
    &(15)  }
$$
Also in this case, the expression of $A^{({\rm d})}$ and $C^{({\rm
d})}$ should be considered to be the lowest order expansions near
$\Delta^{({\rm d})} = 0$.
In fact, in the absence of the dimerization, the exact form of
$\eta^{({\rm d})} \equiv (2\pi)^{-1}\sqrt{C^{({\rm d})} / A^{({\rm
d})}}$ is obtained from the application of the exact solution of the
eight-vertex model as${}^5$
$$
    \eta^{({\rm d})}
    = 2 / [1 + (2/\pi) \sin^{-1} \Delta^{({\rm d})}]~.
    \eqno(16)
$$
If we expand eq.(16) near $\Delta^{({\rm d})} = 0$, we can see that it
agrees with the expression of $\eta^{({\rm d})}$ from eq.(11) and (14).

\bigskip

{}~~~~~~~~$\tilde\cH$ has the same form as $\tilde\cH^{({\rm d})}$ with
the identification of the parameters
$$
    \Delta^{({\rm d})} = {2\Delta_0 \over \sqrt{3}}~,~~~~
    \delta_\perp^{({\rm d})} = {2\delta_\perp \over \sqrt{3}}~,~~~~
    \delta_z^{({\rm d})} = \delta_z~.
    \eqno(18)
$$
Then we can use the knowledge on the dimerized $XXZ$ model.
The phase diagram${}^{6-8}$ of the dimerized $XXZ$ model when
$\,\delta_\perp^{({\rm d})} = \delta_z^{({\rm d})} = \delta^{({\rm
d})}\,$ is shown in Fig.4.
This phase diagram was obtained by the renormalization group
calculation,${}^6$ by the high temperature series expansion after
mapping $\cH^{({\rm d})}$ onto the finite-temperature classical 2D
model (modified Ashkin-Teller model),${}^7$ and by the numerical
diagonalization of the original spin Hamiltonian for finite
systems.${}^8$
When $\delta_\perp^{({\rm d})} = \delta_z^{({\rm d})}$, a naive
consideration leads to the effective dimerization parameter
$$
    \delta^{({\rm d})}_{\rm eff}
    = (2\delta^{({\rm d})}_\perp +  \delta^{({\rm d})}_z) / 3~,
    \eqno(18)
$$
because $\delta^{({\rm d})}_\perp$ is related to $S^x$ and $S^y$,
whereas $\delta^{({\rm d})}_z$ to $S^z$.
However, the renormalization group method and the variational method
bring about${}^4$
$$
    \delta^{({\rm d})}_{\rm eff}
    = (2\delta^{({\rm d})}_\perp + \eta \delta^{({\rm d})}_z) / (2 +
\eta)~.
    \eqno(19)
$$
{}From eqs.(5), (17) and (19), we obtain the mapping of the present
model onto the dimerized $XXZ$ model, as shown in Fig.4.
The case $\gamma = 1$ of the present model corresponds to the case
$\Delta^{({\rm d})} = -2/3\sqrt{3} = -0.385,~\delta^{({\rm d})}
\allowbreak = 0.374$ of the dimerized model, and the case $\gamma =
\infty$ to the case $\Delta^{({\rm d})} = -1,~ \delta^{({\rm d})}
=1/2$.
Therefore there exists the critical value $\gamma_{\rm C}$, where the
transition from the plateau state to the no-plateau state.
This transition is of the Kosterlitz-Thouless type, as known from the
critical properties of the sine-Gordon Hamiltonian.${}^9$

\bigskip

{}~~~~~~~~The above discussion is based on the bosonization method,
which make it difficult to estimate the value of $\gamma_{\rm C}$
itself.
It is because the parameters are slightly shifted through the
bosonization procedure, as already explained.
A rough estimation of the value of $\gamma_{\rm C}$ is
$$
    \gamma_{\rm C} = 4 \sim 5~.
    \eqno(20)
$$

\bigskip

{}~~~~~~~~I have analytically investigated the plateau in the
magnetization curve of the $S=1/2$
ferromagnetic-ferromagnetic-antiferromagnetic spin chain, which is
first pointed out by Hida${}^1$ by the use of the numerical
diagonalization.
The present analytical study semi-qualitatively explains the numerical
result of Hida.

\vfill\eject

\noindent
{\bf References}

\bigskip

\item{1.} K. Hida, {\it J. Phys. Soc. Jpn.} {\bf 63,} 2359 (1994).

\item{2.} Y. Ajiro, T. Asano, T. Inami, H. Aruga-Katori and T. Goto,
{\it J. Phys. Soc. Jpn.} {\bf 63,} 859 (1994).

\item{3.} for instance, E. Fradkin, {\it Field Theories of Condensed
Matter Systems}, Chap.4, Addison Wesley, Redwood City (1991).

\item{4.} K. Okamoto, D. Nishino and Y. Saika, {\it J. Phys. Soc. Jpn.}
{\bf 62,} 2587 (1993).

\item{5.} J. D. Johnson, S. Krinsky and B. McCoy, {\it Phys. Rev. A}
{\bf 8,} 2526 (1973).

\item{6.} K. Okamoto and T. Sugiyama, {\it J. Phys. Soc. Jpn.} {\bf
57,} 1610 (1988).

\item{7.} M. Kohmoto, M. den Nijs and L. P. Kadanoff, {\it Phys. Rev.
B} {\bf 24,} 5229 (1981).

\item{8.} S. Yoshida and K. Okamoto, {\it J. Phys. Soc. Jpn.} {\bf 58,}
4367 (1989).

\item{9.} for instance, J. B. Kogut, {\it Rev. Mod. Phys.} {\bf 51,}
659 (1979).

\vfill\eject
Figure Captions:
\bigskip
\bigskip
Fig.1. Ferromagnetic-ferromagnetic-antiferromagnetic spin chain.
Solid lines and dotted lined represent ferromagnetic couplings $J_{\rm
F}$ and antiferromagnetic couplings $J_{\rm A}$, respectively.
Three spins in an ellipse make an $S = 3/2$ quartet when $\Delta = 1$
and $\gamma \rightarrow \infty$.

\bigskip
Fig.2. Sketch of Hida's numerical result for the magnetization curve.
The case (a) is for smaller $\gamma$ and (b) for larger $\gamma$.

\bigskip
Fig.3. Dispersion relation of $\cH$ of eq.(4) in case of $\Delta = 0$.
(a): $\delta > 0$ case. (b): $\delta = 0$ case.

\bigskip
Fig.4. Mapping of the present model onto the dimerized $XXZ$ chain.
The open circle corresponds to the $\gamma=1$ case and the closed
circle to the $\gamma = \infty$ case.

\end